\documentclass[12pt]{article}

\usepackage{amsmath}
\usepackage{amssymb}
\usepackage{amsfonts}
\usepackage{latexsym}
\usepackage{bm}
\usepackage{color}

\catcode `\@=11 \@addtoreset{equation}{section}

\catcode `\@=12

\newcommand{\be}{\begin{equation}}
\newcommand{\en}{\end{equation}}
\newcommand{\bea}{\begin{eqnarray}}
\newcommand{\ena}{\end{eqnarray}}
\newcommand{\beano}{\begin{eqnarray*}}
\newcommand{\enano}{\end{eqnarray*}}
\newcommand{\bee}{\begin{enumerate}}
\newcommand{\ene}{\end{enumerate}}

\newcommand{\mb}{\mathbb}
\newcommand{\A}{{\mathfrak A}}

\newcommand{\Rc}{{\cal R}}
\newcommand{\mc}{\mathcal}

\newcommand{\1}{{\mb I}}

\newcommand{\Sc}{{\cal S}}

\newcommand{\Hil}{\mc H}

\catcode `\@=11 \@addtoreset{equation}{section}
\catcode `\@=12

\begin{document}

\thispagestyle{empty}

\vspace*{2cm}

\begin{center}
{{\Large \bf Quantum field inspired model of decision making:
Asymptotic stabilization of belief state  via interaction with surrounding  mental environment}}\\[10mm]

{\large Fabio Bagarello} \footnote[1]{ Dipartimento di Energia, Ingegneria dell'Informazione e Modelli Matematici,
Scuola Politecnica, Universit\`a di Palermo, I-90128  Palermo,  INFN, Sezione di Napoli, ITALY and \\ Department of Mathematics and Applied Mathematics, Cape Town, South Africa\\
e-mail: fabio.bagarello@unipa.it\,\,\,\, Home page: www1.unipa.it/fabio.bagarello}

{\large Irina Basieva} \footnote[2]{ Department of Psychology, City University,
London, UK\\
e-mail: Irina.Basieva@gmail.com}

{\large Andrei Khrennikov} \footnote[3]{ International Center for Mathematical Modeling,
in Physics and Cognitive Science\\
Linnaeus University, V\"{a}xj\"{o}, Sweden\\
National Research University of  Information Technologies, Mechanics and Optics (ITMO) \\
St. Petersburg, Russia\\
e-mail: Andrei.Khrennikov@lnu.se}

\end{center}

\vspace*{2cm}

\begin{abstract}
This paper is devoted to justification of quantum-like models of the process of decision making based on the theory of open quantum systems, i.e. decision making is considered as decoherence. This process is modeled as interaction of a decision maker, Alice,  with a mental (information) environment  $\Rc$
surrounding her. Such an interaction generates ``dissipation of uncertainty'' from Alice's belief-state  $\rho(t)$ into  $\Rc$ and
asymptotic  stabilization of  $\rho(t)$  to a steady belief-state. The latter is treated as the decision state. Mathematically the problem under study
is about finding constraints on $\Rc$ guaranteeing such stabilization. We found a partial solution of this problem
(in the form of sufficient conditions). We present the corresponding decision making analysis for one class of mental environments, the so-called
``almost homogeneous environments'', with the illustrative examples: a) behavior of electorate interacting with the mass-media ``reservoir'';
b) consumers' persuasion. We also comment on other classes of mental environments.
\end{abstract}

{\bf keywords:} decision making; quantum-like model; mental (information) environment;  open quantum systems; dissipation of uncertainty; voters' behavior;
consumers' persuasion

\section{Introduction}

The recent years were characterized by explosion  of interest in applications of the mathematical formalism
of quantum theory to studies in cognition, decision making, psychology, economics, finance, and biology,
see, e.g., the monographs \cite{KHR1}-\cite{Asano2015} and a few representative papers \cite{KHR2004}-\cite{Danilov} (the first steps in this direction were
done long  time ago, see,  e.g., \cite{KHR99}).

The approach explored in such mathematical modeling is known as {\it quantum-like.}
In this approach an agent (human, animal, or even cell) is considered as a {\it black box}
processing information in accordance with the laws of quantum information and probability
theories. Thus the quantum-like modeling is basically quantum informational modeling (although this
characteristic feature is typically not emphasized, cf., however, with \cite{INF_BIO}).\footnote{See, e.g.
D' Ariano \cite{R1a} and Plotnitsky \cite{PL3} for the information approach to quantum mechanics.}

 Quantum-like models have to be sharply distinguished from genuinely  quantum physical models of cognition
which are based on consideration of quantum physical processes in the brain, cf. with R. Penrose \cite{P1}
and S. Hameroff \cite{H1}. Although the quantum physical models  have been criticized for mismatching
between the temperature and space-times scales of the quantum physical processes and neuronal processing in the brain, see especially
Tegmark \cite{T1}, they cannot be rejected completely and one may expect that quantum-like models of cognition
 will be (soon or later) coupled with real physical processes in the brain, see \cite{deBarros2009}-\cite{Busemeyer} for some steps in this direction.

The quantum-like approach generated  a variety of models of cognition and decision making.
In the simplest model \cite{KHR99}, \cite{KHR1}, the mental state (the belief state) of  an agent,
Alice, is represented as a quantum state $\psi$ and questions or tasks as quantum observables (Hermitian operators).
Answers to the questions are given with probabilities as determined by Born's rule. For Hermitian operator $A$ with the purely discrete spectrum, Born's rule can be written as
\begin{equation}
\label{BR}
p(A= \alpha_k)= \Vert P_{\alpha_k} \psi \Vert^2 = \langle  P_{\alpha_k} \psi, \psi \rangle,
\end{equation}
where $\alpha_k$ is an eigenvalue of $A$ and $P_{\alpha_k}$ is the projector on the eigenspace corresponding to this eigenvalue.

This model does not describe dynamics of the belief state in the process of decision making. Consideration of dynamics was introduced in the works  of Khrennikov \cite{KHR2004, KHR2006},
 and Pothos and Busemeyer \cite{Pothos and Busemeyer}.
In their dynamical model as in the previous models, an observable  $A$  corresponding to a question (task) faced by Alice is represented as a Hermitian operator. Then Hamiltonian $H$  generating unitary dynamics
\begin{equation}
\label{BRSHR}
\psi(t)= U(t)  \psi_0, \; U(t)= e^{-it H}
\end{equation}
 of the initial belief state $\psi_0$ is introduced, and Alice's decision is represented as measurement of the observable  $A$  at some instant of time.
The authors of \cite{Pothos and Busemeyer} presented cognitive arguments  {supported by experimental studies}
 to determine the instant $t_m$ of measurement. Here the probability of a particular answer is also determined by the Born rule, but applied to belief state $\psi(t_m).$
{ Of course, this is an important issue, since different values ot $t_m$ can give rise to completely different results.
 In spite of the partial progress in determination of $t_m$ in the article of Pothos and Busemeyer \cite{Pothos and Busemeyer}, this complex problem cannot be considered completely solved.
 Moreover, in solving this problem Pothos and Busemeyer had to go beyond the quantum theory and to appeal to psychological theoretical and experimental studies. It would be attractive
 to solve this problem entirely in the quantum framework. }

{We remark that the Schr\"odinger equation  describes  the  dynamics of an isolated system. In the presence of an environment, the dynamics of the system is non-unitary. Approximately (under some
sufficiently natural conditions) it is described by the Gorini-Kossakowski-Sudarshan-Lindblad (GKSL) equation (often called simply the Lindblad
equation), the simplest version of the quantum master equation.  One of the main distinguishing features of such dynamics
is that it does not preserve the pure state structure: it (immediately) transforms
a pure initial state $\psi_0$ into a mixed state given by the density operator:
\begin{equation}
\label{BRSHR1}
\rho(t)=  U(t)  \rho_0, \; U(t)= e^{-it L},
\end{equation}
 where $L$ is the generator of the GKSL-evolution and $\rho_0=\vert  \psi_0\rangle \langle \psi_0\vert.$ This dynamics is in general non-unitary.}
 {This equation describes the process of {\it the system adaptation to the surrounding environment.} This is the complex dynamical process combining the internal
 state dynamics of the system with adaptation to signals received from the environment. If the dynamics is discrete with respect to time, then it can be represented
as {\it a chain of unitary evolutions and (generalized)  quantum Bayesian updates.}}\footnote{{As was quickly understood in quantum physics, the L\"uders projection postulate
describes only one very special class of the quantum state updates resulting from measurements. We remark that already von Neumann accepted applicability
of this straightforward form of the state update only for observables with non-degenerate spectra. Generally, in the case of an observable with degenerate spectrum, a pure pre-measurement state
can be transferred into a mixed state \cite{VN}. Later these considerations of von Neumann were elaborated in the form of
the {\it theory of quantum instruments}, see \cite{KHR_HB_QI} for non-physicist friendly presentation. The most consistent justification
of this theory is obtained in the framework of the {\it theory of  open quantum systems}.}} { In this paper we cannot discuss this interesting issue in more detail, see \cite{ASANO_EPI}
for detailed consideration of a two dimensional example with application to the evolution theory.}

{In fact, GKSL-dynamics does not contradict the Schr\"odinger equation structure of the quantum evolution. Let us denote the system under study by $S$ and the surrounding
environment by ${\cal R}$ (``reservoir'' for $S.)$ Suppose that initial state of  the compound system $S+{\cal R}$ is pure and separable. Dynamics in the state space of
$S+{\cal R}$ is still unitary and given by a Hamiltonian for $S+{\cal R}.$ The main distinguishing feature of this unitary dynamics is that (in the presence of interaction
between $S$ and ${\cal R})$ it induces {\it entanglement} and the state of the compound system becomes not more separable.
Hamiltonians for the composite system $S+{\cal R}$  are very complex, since they includes, in general, an infinite number of degrees of freedom
of ${\cal R}.$ Typically it is impossible to solve the Schr\"odinger equation for the state of composite system $S+{\cal R}.$ (Although in some special cases, as those considered in this paper and others discussed in \cite{bagbook}
analytic solutions can be found.) Therefore, most studies are restricted to the dynamics of the state $\rho(t)$ of $S$ alone, which is described (approximately) by the GKSL-equation.
But even if one were able to solve the Schr\"odinger equation for  $S+{\cal R},$ the solution would be a very complex infinite-dimensional state vector. Since we are interested in behavior of
$S,$  we would then take the trace with respect to all degrees of freedom of   ${\cal R}$ and obtain the state of $S.$ A simple mathematical theorem implies that in presence of entanglement
this trace-state cannot be pure, i.e., the state is descirbed by density operator.  Its dynamics under the GKSL-equation is known as {\it decoherence:}  decreasing of state's purity (or coherence)
in the process of interaction with an environment.}

{Since consideration of an isolated cognitive system is even a higher degree idealization than consideration of an isolated physical system, it is natural to modify the dynamical scheme of decision making based on unitary
Schr\"odinger dynamics \cite{KHR2004, KHR2006, Pothos and Busemeyer} and consider general dynamics of the belief-state, either by using the approximative GKSL-dynamics
 or by tracing the state of the compound system.
Roughly speaking there is no choice: either one has to ignore the presence of environment or consider non-unitary dynamics, e.g., (\ref{BRSHR1}).  Of course, such non-unitary
dynamical model of decision making is much more mathematically complicated. However, it has one very important advantage. Here the aforementioned problem  of determination
of $t_m$ is solved automatically, although not straightforwardly.}

For a natural class of quantum master equations, in the limit $t \to \infty$ the system's state $\rho(t)$ approaches a {\it  steady state} $\rho_{\rm{out}}.$
Diagonal elements of this operator (with respect to  the ``pointer basis'' corresponding
to the observable $A$) under measurement give the probabilities of measurement outputs.
The model of decision making as decoherence was represented in a series of papers of Asano et al., see, e.g.,  \cite{Asano2011}, \cite{Asano2011a}
\cite{Asano2012a}, \cite{Asano2015}.
The model is purely informational.  Both a ``quantum-like system", Alice,  and  her mental (or information) environment (``bath'',
reservoir'')  are represented by   quantum states, $\psi$ and $\phi.$ (It can be assumed that initially these states are pure.)
We are not interested in their physical or neurophysiologic   realizations.
Asano et al.  \cite{Asano2011} applied the {\it theory of open quantum systems} and the GKSL-equation
to model experimental data collected in decision making experiments. The initial belief state is typically represented as a pure state of complete uncertainty,
\begin{equation}
\label{BR1}
\psi=(\vert 0\rangle+ \vert 1\rangle)/\sqrt{2},
\end{equation}
where $\vert i \rangle, i=0,1,$ are eigenstates of the dichotomous observable $A$ representing a ``yes''/``no'' question.
A proper decision making dynamics should asymptotically ($t \to \infty)$ drive the belief state of Alice $\rho(t)$  to the  output (mixed) state
 represented by the density operator $\rho_{\rm{out}}.$ Its diagonal elements in the basis of eigenvectors of the operator $A$ give probabilities of possible decisions.

We remark that the open quantum systems approach to decision making can be considered as a possible realization of the contextual treatment of cognition, cf.
Khrennikov \cite{KHR1}, \cite{KHR2} and Dzhafarov et. al. \cite{24}, \cite{Dzhafarov2014}. The surrounding mental environment represents a measurement context for
decision making.

{Of course, Alice cannot wait for $t\to \infty$ to make a decision. And, as we know in physics, a quantum system relaxes to a steady state very quickly. Here we have to
point to difference between a mathematical model and its applications to real phenomena. The notion of  limit is a mathematical abstraction. Of course, it is not applicable
to real physical or cognitive systems. In practice, the limit procedure encodes the process of approaching some quantity. The existence of a limit for the density operator $\rho(t)$
guarantees that its fluctuations  decrease. If the magnitude of fluctuations becomes smaller than some $\epsilon>0,$ then such $\rho(t_\epsilon)$ can be selected as a good
approximation of the steady state (the latter is also a mathematical abstraction, in real physical processes it is never approached exactly).  This $t_\epsilon$ plays the role of
$t_m$ from the unitary dynamical model of decision making. Moreover, more complicated considerations based on the theory of decoherence demonstrate that this instant of time
$t_\epsilon$ can be identified with the {\it relaxation time} $T.$ The latter is determined by the structure of interactions between a system and its environment.
 Typically in quantum physics the time interval $[0, T]$ is very short and a system relaxes very quickly to its steady state. Since such considerations would make the paper even more complex mathematically,
we shall not proceed in this direction. (See, e.g., \cite{RT} for derivation of an analytical expression for the relaxation time as a function of the heat-bath and interaction parameters.)
The main message of these considerations to a reader is that  $\lim_{t \to \infty}$ is just
an abstraction (at $t= \infty$ all fluctuations disappear completely). In reality a system relaxes very quickly to its steady state.  }

The main problem of the quantum(-like) decision theory is to construct an operator representation of  Hamiltonians and ``Lindbladians''
(the latter represents interactions of Alice with her mental reservoir).
In contrast to quantum physics, there is no analog of the classical phase  formalism for cognition and decision making, i.e., we cannot use the quantization procedure to transform
functions on the classical phase space into operators - the procedure of Schr\"odinger's quantization is not applicable.

In quantum theory one can also use the quantization procedure based on  the algebra of operators of creation and annihilation. It is especially useful for
second quantization (quantum field theory), see appendix 1 for brief presentation of the basics.  This sort of quantization can be successfully applied to quantum-like modeling. Consider again a dichotomous  question $A,$
 with the values ``no''/ ``yes''. Creation operator $a^\star$ creates  ``yes'' from ``no'', annihilation operator $a$  transforms
``yes'' to ``no''.  We can compound Hamiltonians and observables with the aid of creation and annihilation operators (similarly, e.g., to quantum optics) .
This approach  to construction of operators for  problems of decision making was pioneered by Bagarello  \cite{bagbook} and coauthors  who applied it, see, e.g.,
\cite{Bagarello2015a}, \cite{Bagarello2015b}, \cite{Bagarello2016}, to a variety of problems.
Bagarello et al. were interested in time dynamics of averages; in particular, probabilities were treated
statistically. In  series of works \cite{Bagarello2015a}, \cite{Bagarello2015b}, \cite{Bagarello2016}, a system was interpreted as an
agent and the reservoir had a purely information interpretation, see especially \cite{BHKH} for a discussion.

The same approach based on creation-annihilation operators can be explored to model decision making at  individual level (with the
subjective interpretation of probabilities). Here one can explore the scheme which was approved in the works of Asano et al. \cite{Asano2015}: consideration of the asymptotic
dynamics of the belief state and using the asymptotically output state (for $t\to \infty)$ as the basis of  decision making. The first step in this direction was done in paper
\cite{Asano2011}. The aforementioned decision making scheme
(based on asymptotic stabilization of the mental (belief) state of an agent)  generates interest to study the asymptotic dynamics of the system interacting with a reservoir.
This is a nontrivial mathematical problem and its partial solution (for specially designed information reservoirs) is presented in this paper.

In section \ref{DM7} we discuss several aspects of  the process of decision making for a dichotomous question $A$ through belief-state stabilization. Then, in section \ref{sect_SP}, the mathematical problem is discussed in more details. More explicitly, in section \ref{sectAahr} we prove  stabilization of the
 belief state in the case of the ``almost homogeneous mental environment'' of Alice. We start
with the case of a  question' $A$ having an infinite number of possible outcomes labeled as $n_a=0,1,..., n,...\;.$ Then in section \ref{FREM} we consider the
case of a ``dichotomous question'' $A$ having two possible outcomes, $n_a=0,1.$ In section \ref{CR7} we generalize our theory to the case of environments having a more complex
structure.  Section \ref{conclrem} contains our conclusions, while
{appendices 1 and 2 are devoted to few introductory remarks on canonical anti-commutation relations (CAR) and some extra mathematical details.}

\section{Decision making as decoherence}
\label{DM7}

Consider the case of  two dichotomous questions posed to Alice, $A=0,1$ (``no'', ``yes'') and the mental environment  $\Rc$ similarly composed
of  dichotomous degrees of freedom. In  section \ref{FREM} this situation will be modeled by using operators $a, a^\dagger$ (Alice's operators)
and $b, b^\dagger$ ($\Rc$'s operators) satisfying the CAR:
\begin{equation}
\label{LIS}
\{a, a^\dagger\}=\1, \{b, b^\dagger\}=\1, \{a, b\} =0,
\end{equation}
where the anti-commutator of two operators $x,y$ is defined as  $\{x,y\}= xy+yx.$ Moreover, $a^2=b^2=0$.
In quantum field theory, see appendix 1,  these operators are known as {\it the operators of creation and annihilation.}\footnote{{In physics these operators represent the processes of creation
and annihilation of fermions, e.g., electrons.}}
The operators related to $\Rc$ can depend on some parameter $k$ (discrete or continuous) representing degrees of freedom of $\Rc, b=b(k), b^\dagger= b^\dagger(k).$
Typically environment $\Rc$ has a huge number of degrees of freedom and belongs to infinite-dimensional Hilbert state space ${\cal K}.$
In the case of dichotomous questions asked to Alice, it can be assumed that her state space ${\cal H}$ has the dimension of two, {\it the qubit space of quantum information theory.}

 In quantum field theory the terminology ``creation-annihilation operators''  has coupling to real physical systems. The operators represent the processes of
creation-annihilation of quantum particles (e.g., photons or electrons), see appendix 1.
As was emphasized in introduction, our model is of the purely  informational nature. Therefore the operators $a, a^\dagger$ and  $b, b^\dagger$ have to be treated as the
formal representation of the processes of ``creation'' and  ``annihilation'' of information states.\footnote{In principle, we can simply call these operators
ladder operators as is done in formal mathematical theory.}

{Consider, e.g., Alice's operators $a, a^\dagger.$ Let
$\phi_0=\vert 0\rangle, \phi_1=\vert 1\rangle$ be the orthonormal basis in Alice's state space ${\cal H}$ corresponding to the answers ``no''/``yes'' to the question $A.$
As a consequence of the CAR, the operators $a, a^\dagger$ act on these basis vectors in the following way:
$a^\dagger \vert 0\rangle= \vert 1\rangle,  a^\dagger \vert 1\rangle= 0;  a \vert 1\rangle= \vert 0\rangle, a \vert 0\rangle=0.$}

{The operators  $a$ and $a^{\dagger }$ modify Alice's attitude to selection of alternatives.  In this framework
these operators are considered as \emph{reflection operators} \cite{BHKH}.  We represent the question $A$ as the {\it number operator} of Alice,
$\hat{n}_a=a^{\dagger } a.$  The eigenvalues of this operator, $n_a=0,1,$ correspond to the choices of Alice at $t=0.$
Therefore it is natural to name $ \hat{n}_a$ the {\it decision operator.}\footnote{In quantum field theory the number operator has the meaning of the
number of physical particles, see appendix 1.  In our model it represents states indexed by numbers.}}

{Consider the simplest situation which can be modeled in this framework: Alice is asked a question $A$ and she is  surrounded  by a population $\Rc$
whose members are asked the same question $A.$ Alice interacts with this population  $\Rc$ and she gets to know behavior of its members with respect to this question.  Of course, she cannot ``scan''
$\Rc$ completely to know the concrete answers of people to $A.$ She just gets to know a sort of average $n_b$ of answers to $A.$ In the simplest case of a homogeneous
population this average is a constant.  In the general case average $n_b$ can non-trivially depend on the parameter $k$ encoding various
population clusters $\Rc(k): \; n_b=n_b(k).$  Here Alice (through interaction with this population) obtains information about averages  for the  answers to $A$
corresponding to different clusters $\Rc(k)$  of  $\Rc.$}

{In principle, $\Rc$ needs not be combined of physical agents. As was emphasized in introduction, our model is of the purely informational nature. The environment
$\Rc$ can, for example, represent mass-media's  image of the question under consideration: the image created in TV-debates and shows, web-blogs, newspapers. For example,
$A$ can be a referendum question: {\it ``vote for Brexit or against?''} (or the recent USA-election question:
{\it ``vote for Trump or not?''}). By analysing (generally unconsciously)  mass-media's opinions Alice estimates the average
$n_b$  (or in general the averages $n_b(k))$ and she makes her decision based on these averagess.}

{In a more general situation  the members of $\Rc$ express their attitude with respect to a variety of dichotomous questions and the $A,$ the question asked to Alice, can be among them,
but not necessarily. In the latter case Alice makes her decision based on behavior of the surrounding environment  $\Rc$ (physical or informational) without even  understanding that her decision is made through influence of $\Rc.$}

Now we consider dynamics of decision making. In section \ref{sect_SP} we shall use the Heisenberg picture, i.e., we are concerned with the dynamics of operators. We are interested in the dynamics
of the reflection operators, $a^\dagger(t), a(t),$ with initial conditions $a^\dagger(0)=a^\dagger, a(0)=a.$
Based on the reflection operators dynamics we can find the
the decision operator dynamics  $\hat n_a(t)=a^\dagger(t) a(t)$ and, hence, its average
\begin{equation}
\label{za}
n_a(t):=<\hat n_a(t)\otimes \1>=<a^\dagger(t)a(t)\otimes \1>
\end{equation}
with respect to some initial state of the compound system,  see section \ref{sect_SP} for details.

In the case of dichotomous questions  the average   $n_a(t)$ has the straightforward probabilistic
meaning. To see this, let us expand expression (\ref{za}). Denote the state space  of
the compound system, Alice and her mental environment,  by the symbol ${\cal H}\otimes {\cal K}$ and the
operator of  Heisenberg evolution of this system by $U_t.$
For an arbitrary initial state  $\left<.,.\right>_R$,  formula (\ref{za}) reads as
$$n_a(t)=
< U_t (a^\dagger \otimes \1) U_t^\dagger U_t (a\otimes \1) U_t^\dagger >_R=\rm{Tr}  U_t (a^\dagger a\otimes \1) U_t^\dagger R.
$$
By using the cyclic property of the trace we get:
$$
n_a(t)= \rm{Tr}  (a^\dagger a\otimes \1) U_t^\dagger R U_t =  \rm{Tr} R(t) (a^\dagger a\otimes \1) =
 <a^\dagger a\otimes \1>_{R(t)},
$$
where $R(t)= U_t^\dagger R U_t$ represents the state dynamics (so we transferred the operator dynamics into the the state dynamics,
from the Heisenberg picture to the Sch\"odinger picture).\footnote{The Schr\"odinger picture providing the belief-state dynamics is basic
for the probabilistic interpretation. However, equations in the Heisenberg picture are easier for analytic treatment. Therefore in section \ref{sect_SP}
we shall work in the Heisenberg picture. }
Consider now the partial trace of $R(t)$ with respect to the environmental degrees of freedom,
$\rho(t)= \rm{Tr}_{{\cal K}} R(t).$  It represents the dynamics of Alice's belief-state in the process of her interaction with the environment.
Consider the average of the decision operator $\hat n_a=  a^\dagger a =a^\dagger (0) a(0)$ with respect to this belief-state:
$$
< \hat n_a>_{\rho(t)}= \rm{Tr} \rho(t)    \hat n_a.
$$
This average is directly related to probability. Since the decision operator $\hat n_a$ has two eigenvalues, $n_a=0,1,$
the average coincides with the probability:
\begin{equation}
\label{LBq0}
P_t(A=1)=    < \hat n_a>_{\rho(t)}= <\rho(t) \phi_{1},\phi_{1}>   .
\end{equation}
We recall that here the question $A$ is represented by the operator  $\hat n_a.$ The eigenvector $\phi_{1}$ corresponds to the answer
``yes'' to this question. $P_t(A=1)$ is the subjective probability in favor of this answer which Alice assigns at the moment of time $t.$
We remark that this assignment can be done unconsciously. Alice will consciously report only her final decision, see considerations below.

Now we make a very general remark about the partial trace. For any operator $M$ acting in ${\cal H},$ the following equality holds:
$$
\rm{Tr} \rho M = \rm{Tr} R (M\otimes \1), \; \mbox{where}\; \rho=  \rm{Tr}_{{\cal K}} R.
$$
Therefore the average $n_a(t)$  equals to the average of Alice's decision operator:
\begin{equation}
\label{LBq}
n_a(t)= < \hat n_a>_{\rho(t)}.
\end{equation}
In section \ref{sect_SP} we obtain the averages of the decision operator $\hat n_a(t)$ for its  eigenstates $\phi_{n_a}$ (for operators satisfying CAR, we have: $n_a=0,1).$ Thus initially
Alice has a definite belief about possible answers to the question $A.$

In section \ref{sectAahr} we consider the case of the ``almost homogeneous reservoir'' - a mental (information) environment  $\Rc$ which is behaviorally almost homogeneous in the following sense (see section \ref{sectAahr} for
the mathematical formalization). The basic parameter characterizing  $\Rc$ is deduced by $<b^\dagger(k)b(q)>= n_b(k) \delta(q-k),$ the average for $k$-type agents. For an almost
homogeneous  $\Rc$, $n_b(k)$ is constant. In general, however, it can vary with $k$ \footnote{In the simplest
model all agents in $\Rc$ are asked the same question $A$ as Alice and $n_b(k)$ is the average of their answers to $A.$ In a more complex environment $\Rc$ each cluster $\Rc(k)$
of agents labeled by the parameter $k$ is asked its own question $A(k)$ and $n_b(k)$ is the average of their answers to $A(k).$}.
It must be stressed that the reservoir considered in section \ref{sectAahr} is not completely homogeneous.
Its members are not all identical. They have different internal dynamical scales represented by the function  $\omega(k)= \omega_b k$  which  really depends on $k$,
even if this dependence is {\em relatively mild} ($\omega(k)$ is linear in $k$, so that $\omega(k_1)\simeq\omega(k_2)$ if $k_1\simeq k_2$), see again section \ref{sectAahr}.

For such environment $\Rc,$ the average $n_a(t)$ of the decision operator $\hat n_a$ was found, see formula (\ref{sr7}) section \ref{sectAahr} (with the corresponding generalization
to the  CAR-case in section \ref{FREM}).  This formula in combination with equalities (\ref{LBq0}), (\ref{LBq}) gives the probabilistic dynamics:
\begin{equation}
\label{LB}
P_t(A=1) = n_a(t)=  n_a\,e^{-\frac{2\lambda^2\pi}{\omega_b} t}+n_b\left(1-e^{\frac{2\lambda^2\pi}{\omega_b} t}\right)
= n_b + (n_a - n_b) \,e^{-\frac{2\lambda^2\pi}{\omega_b} t},
\end{equation}
where $\lambda$ is the constant of interaction  between Alice and $\Rc,$ see (\ref{sr4}). { As
we will see in the next section, this is the result of a reasonable choice of the Hamiltonian operator, where the interactions between
Alice and her reservoir are included}\footnote{This approach, which is used everywhere in \cite{bagbook}, is the standard way in which the dynamics
of any system, micro- or macroscopic, is deduced in classical and in quantum mechanics. What we are doing, in fact, is to adapt the same basic idea to decision making processes.}.

Consider, e.g., the case $n_a=1,$ e.g., initially Alice was for Brexit. Suppose that her surrounding environment is almost homogeneous\footnote{The latter assumption  is not
so unusual. Although the modern information (mental) environment represents huge variety of information flows, an individual is typically coupled to the concrete flow, e.g.,
she has the custom to see only particular TV-channels and follow only specially selected Internet resources. Even her physical human
environment is typically homogeneous. A professor of a university is surrounded by rather homogeneous population with liberal views, including Brexit, Trump or Putin.
One of the authors of this paper was visiting USA just before the second Bush-vote. The university environment expressed generally the belief that Bush would not be elected. For such $\Rc,$ the average $n_b$ could be estimated as approximately equal to one.} and $n_b=1/3,$ i.e.,
its members express sufficiently strong anti-Brexit attitude. Then Alice following the dynamics expressed by (\ref{LB}) would lose her inclination to vote for Brexit and finally she may
accept the surrounding attitude to vote against Brexit.
{It is interesting to notice that the rapidity of the decision process is directly related to the strength of the interaction between
 Alice and her environment, and inversely connected to $\omega_b$. This is in agreement with the fact, see \cite{bagbook}, that $\omega_b$ describes a sort of inertia of the system.}

The ``decision probability'' is given by the expression:
\begin{equation}
\label{LB1}
P(A=1)= \lim_{t\to \infty} P_t(A=1) =   n_b.
\end{equation}
We remark that dependence on the initial belief-state of Alice disappeared. It does not matter whether she was in the belief-state $\phi_{0}$ (she was firmly determined to reply ``no'')
or in the belief-state $\phi_{1}$ (she was firmly determined to reply ``yes''). Finally, she set the {\it subjective probability}   $P(A=1)= n_b$ to reply ``yes''. Moreover, as we will discuss later,
even if initially Alice was in a superposition belief-state (i.e., she started with a belief-state of uncertainty), it would lead to setting of the same probability in favor to reply ``yes''.
This behavior of Alice is natural: she interacts with a homogeneous reservoir, an ensemble of agents, where all agents have the same probability to favor the ``yes'' answer;
Alice simply follows them.

In spite of its simplicity, the belief state dynamics of a decision maker interacting with an almost homogeneous information environment can have interesting applications.
It describes well not only modern mass-media campaigns on the political arena, but can also  be used in economics and finance.

For example, in this framework we can model the process of {\it persuasion of consumers.} A company wants to persuade consumers  to buy some commodity $C.$
So the question  $A$ is {\it ``to buy or not to buy?''} To approach the goal, the company creates an  information environment  $\Rc$ (almost homogeneous) with the parameter $n_b\approx 1.$
\footnote{Typically this is done through
a massive campaign in the mass-media. However, the information content of  $\Rc$ is not reduced to direct advertising of $C,$ see considerations below.}
Then our model shows that, for consumers,  it is possible to approach the same level of confidence to buy    $C$ as in the surrounding information environment $\Rc.$ As was already remarked,
the information structure of $\Rc$ need not include directly the question  $A,$ i.e., the information campaign need not be straightforwardly oriented to
advertising of the commodity $C.$  The $\Rc$  can be concerned with other questions, probably quite different from $A.$
Then the main task of the company driving the persuasion campaign is to establish interaction of consumers with  $\Rc$ which can be formally represented by the interaction Hamiltonian
considered in this paper.

Persuasion of consumers need not be reduced to the concrete commodity $C.$ A group of corporations can perform persuasion regarding a class of commodities, e.g.,
a new generation of mobile phones, or wind energy, or electric cars. The strategy is the same: creation of an information environment and coupling it with people's opinion in such a way that would generate stabilization of  subjective probabilities $P_t(A=1)$ (in a long run  campaign) to the desired value of $n_b.$

\section{The stabilization procedure}
\label{sect_SP}

This section is devoted to justification of the model of decision making as decoherence, stabilization of the belief-state $\rho(t)$ of Alice.
In the Heisenberg picture this problem can be transformed into the problem of stabilization of reflection (creation-annihilation) operators
$a(t), a^\dagger (t)$ and consequently the decision operator  $\hat n_a(t)= a^\dagger(t) a(t)$ (the number operator). Of course, such stabilization
is possible only for special classes of information (mental) environments and interactions between decision maker and surrounding environment.
Mathematically the problem is very complicated and we were able to solve it only partially.  This section is of the purely mathematical (and quantum physical)
nature.

If $\Sc$ is a closed quantum system driven by a time-independent, self-adjoint, hamiltonian $H$, it is natural to suspect that only periodic or
quasi-periodic effects  can take place, during the time evolution of $\Sc$. This is because the energy of $\Sc$ is preserved, and this seems to
prevent any damping effect in $\Sc$. For instance, if we work in the Schr\"odinger representation (SR), the time evolution  $\Psi(t)$ of the
wave-function of the system is simply $\Psi(t)=e^{-iHt}\Psi(0)$ and, since the operator $e^{-iHt}$ is unitary, we do not expect $\Psi(t)$
to decrease (in some suitable sense) to zero when $t$ diverges. Nevertheless, we will show that a similar decay feature is possible if $\Sc$ is
coupled to a reservoir $\Rc$, but only if $\Rc$ is rather {\em large} as compared to $\Sc$, in particular, if $\Rc$ has an infinite number of
degrees of freedom.

To see this in details, we start by considering a first system, $\Sc$, interacting with a second system, $\tilde\Sc$, and we assume for the time being that both $\Sc$ and
$\tilde\Sc$ are {\em of the same size}: to be concrete,  this means here that $\Sc$ describes a single particle  and, analogously, $\tilde\Sc$ describes a second particle.
To model dynamics, we introduce the operators
$a$, $a^\dagger$  and $\hat n_a= a^\dagger a$ (for the first particle)  and the operators
 $b$,$b^\dagger$ and $\hat n_b= b^\dagger b$ (for the second particle).\footnote{Cognitive meaning of these operators in the quantum-like model of decision
making will be discussed in section \ref{DM7}. }
These operators obey the following {\it canonical commutation relations} (CCR):
$$
[a,a^\dagger]=[b,b^\dagger]=\1,$$
while all the other commutators are assumed to be zero.

A natural choice for a Hamiltonian of  the compound system $\Sc\cup\tilde\Sc$ is the following: $$h=\omega_a\hat
n_a+\omega_b\hat n_b+\mu\left(a^\dagger b+b^\dagger a\right),$$ where $\omega_a, \omega_b$ and $\mu$ must be real quantities for $h$ to be
self-adjoint. Recall that losing self-adjointness of $h$ would produce a non unitary time evolution, and this is out of the scheme usually
considered in ordinary quantum mechanics. The hamiltonian $h$ contains a free part plus an interaction part. The latter is
such that, if the eigenvalue of $\hat n_a$ increases by one unit (during time evolution), then the eigenvalue of $\hat n_b$ decreases by
one unit, and viceversa. This is because $[h,\hat n_a+\hat n_b]=0$, so that $\hat n_a+\hat n_b$ is an integral of motion.

Let us now consider dynamics of operators in the Heisenberg representation.
The equations of motion for $a(t)$ and $b(t)$,  can be easily deduced and turn out to be
$$
\dot a(t)=i[h,a(t)]=-i\omega_a a(t)-i\mu b(t), \qquad a(0)=a,
$$
$$
 \dot b(t)=i[h,b(t)]=-i\omega_b b(t)-i\mu a(t), \qquad b(0)=b.
$$
The solution can be written as $a(t)=a\alpha_a(t)+b\alpha_b(t)$ and $b(t)=a\beta_a(t)+b\beta_b(t)$, where the functions $\alpha_j(t)$ and
$\beta_j(t)$, $j=a,b$, are linear combinations of $e^{\lambda_\pm t}$, with
$\lambda_\pm=\frac{-i}{2}(\omega_a+\omega_b-\sqrt{(\omega_a-\omega_b)^2+4\mu^2})$. Moreover $\alpha_a(0)=\beta_b(0)=1$ and
$\alpha_b(0)=\beta_a(0)=0$, in order to have $a(0)=a$ and $b(0)=b$. Hence we see that both $a(t)$ and $b(t)$, and,as a consequence, $\hat
n_a(t)=a^\dagger(t)a(t)$ and $\hat n_b(t)=b^\dagger(t)b(t)$, are linear combinations of oscillating functions, so that no
damping is possible within this simple model.

\vspace{3mm}

Suppose now that the system $\tilde\Sc$ is replaced by an (infinitely extended) reservoir $\Rc$, whose particles are described by an infinite
set of CCR-operators $b(k), b^\dagger(k)$ and $\hat n(k)=b^\dagger(k) b(k)$, $k\in\Bbb{R}$. { Each $k$ labels one of the elements of the reservoir surrounding $\cal S$. Hence, for $k_1\neq k_2$, we are considering two different elements which may have different characteristics. This is why, in (\ref{sr4}), we are introducing two $k$-depending functions, $\omega(k)$ and $f(k)$. } The hamiltonian of $\Sc\cup\Rc$ extends $h$
above and is now taken to be \be H=H_0+\lambda H_I,\quad H_0=\omega \hat n_a+\int_{\Bbb{R}}\omega(k)\hat n(k)\,dk,\quad
H_I=\int_{\Bbb{R}}\left(a b^\dagger(k)+a^\dagger b(k)\right)f(k) dk, \label{sr4}\en where $[a,a^\dagger]=\1$,
$[b(k),b^\dagger(q)]=\1\delta(k-q)$, while all the other commutators are zero. {It could be useful to notice that here, rather than the CAR used in section 2,
we are assuming CCR. We will see later, in section 3.3, that this does not affect our main conclusions. The reason for the different choices is to show that
the output of the model does not really depend on the commutation rules adopted.} All the constants appearing in (\ref{sr4}), as well as the
regularizing function $f(k)$, are real, so that $H=H^\dagger$.  Notice that an integral of
motion  exists also for $\Sc\cup\Rc$, $\hat n_a+\int_{\Bbb{R}}\hat n(k)\,dk$, which extends the one for $\Sc\cup\tilde\Sc$, $\hat n_a+\hat n_b$.
With this choice of $H$, the Heisenberg equations of motion are \be \left\{
\begin{array}{ll}
\dot a(t)=i[H,a(t)]=-i\omega a(t)-i\lambda \int_{\Bbb{R}} f(k)\,b(k,t)\,dk,\\
\dot b(k,t)=i[H,b(k,t)]=-i\omega(k) b(k,t)-i\lambda f(k)\,a(t),
\end{array}
\right. \label{sr5} \en They are supplemented by the initial conditions $a(0)=a$ and $b(k,0)=b(k)$. In particular, the last equation can be rewritten in an integral form as
\be b(k,t)=b(k)e^{-i\omega(k)t}-i\lambda f(k)\int_0^t a(t_1)e^{-i\omega(k)(t-t_1)}\,dt_1.\label{11}\en

\subsection{An almost homogeneous reservoir}
\label{sectAahr}

In this section we fix $f(k)=1$ and $\omega(k)=\omega_b k$, where $\omega_b\in\Bbb R_+$. This is a standard choice in quantum optics, \cite{barrad}.
We now insert $b(k,t)$ in (\ref{11}) in the first equation in (\ref{sr5}), change the order of integration, and use the integral expression for the Dirac delta
$\int_{\Bbb{R}}e^{-i\omega_b k(t-t_1)}\,dk=\frac{2\pi}{\omega_b}\delta(t-t_1)$, as well as the equality $\int_0^t g(t_1)\delta(t-t_1)\,dt_1=\frac{1}{2}g(t)$ for any test function
$g(t)$. Then, we conclude that
\be
\dot a(t)=-\left(i\omega+\frac{\pi\lambda^2}{\omega_b}\right) a(t)-i\lambda \int_{\Bbb{R}} b(k)\,e^{-i\omega_bkt}\,dk.
\label{12}\en
This equation can be solved, and the solution can be written as \be a(t)=\left(a-i\lambda
\int_{\Bbb{R}}dk\,\eta(k,t)b(k)\right)e^{-\left(i\omega+\frac{\pi\lambda^2}{\omega_b}\right)t}, \label{sr6}\en where $\eta(k,t)=\frac{1}{\rho(k)}\left(e^{\rho(k)t}-1\right)$
and $\rho(k)=i(\omega-\omega_bk)+\frac{\pi\lambda^2}{\omega_b}$. Using complex contour integration it is possible to check, see appendix 2, that $[a(t),a^\dagger(t)]=\1$ for all $t$:
this means that natural decay of $a(t)$, described in (\ref{sr6}), is balanced by the reservoir contribution. This feature is crucial since
it is a measure of the fact that the time evolution is unitarily implemented in our approach, even if $a(t)$ apparently decays for  increasing $t$.

Let us now consider a state over $\Sc\cup\Rc$, $\left<\,X_\Sc\otimes X_\Rc\right>=\left<\varphi_{n_a},X_\Sc\varphi_{n_a}\right>\,\left<\,
X_\Rc\right>_\Rc$, in which $\varphi_{n_a}$ is an eigenstate of the number operator $\hat n_a$ and $<\,>_\Rc$ is a state of the reservoir, which is assumed to satisfy, among other properties
(see \cite{barrad,bagbook}),
\begin{equation}
\label{NB}
\left<\, b^\dagger(k)b(q)\right>_\Rc=n_b(k)\delta(k-q).
\end{equation}
This is a standard choice, see for instance \cite{barrad}, which extends the choice we  made for $\Sc$. Here $X_\Sc\otimes X_\Rc$ is the tensor product of an operator of the
system, $X_\Sc$, and an operator of the reservoir, $X_\Rc$. Then, if for simplicity we take the function $n_b(k)$ to be constant in $k$ we get,
calling $n_a(t):=<\hat n_a(t)>=<a^\dagger(t)a(t)>$, \be n_a(t)=n_a\,e^{-\frac{2\lambda^2\pi}{\omega_b} t}+n_b\left(1-e^{\frac{2\lambda^2\pi}{\omega_b} t}\right), \label{sr7}\en
which goes to $n_b$ as $t\rightarrow\infty$. Hence, if $0\leq n_b<n_a$, the value of $n_a(t)$ decreases with time. If, on the other hand,
$n_b>n_a$, then the value of $n_a(t)$ increases for large $t$. This is the exponential rule which, as discussed before, cannot be deduced if
$\Rc$ has not an infinite number of degrees of freedom. Notice that, in particular, if the reservoir is originally empty, $n_b=0$, then
$n_a(t)=n_a\,e^{-\frac{2\lambda^2\pi}{\omega_b} t}$ decreases exponentially to zero: the system becomes empty. On the other hand, since $\hat n_a+\int_{\Bbb
R}\hat n(k)\,dk$ is a constant of motion, the reservoir starts to be filled up.

\vspace{2mm}

{\bf Remark 1.} The {\em continuous} reservoir considered here ($k\in\Bbb R$) could be replaced by a discrete one, describing again an
infinite number of particles, but  labeled by a discrete index. In this case, to obtain a Dirac delta distribution, which is the crucial
ingredient in the derivation above, we have to replace  the integral $\int_{\Bbb{R}}e^{-ik(t-t_1)}\,dk=2\pi\delta(t-t_1)$ with the
{\em Poisson summation formula}, which we  write here as
$\sum_{n\in\Bbb{Z}}e^{inxc}=\frac{2\pi}{|c|}\sum_{n\in\Bbb{Z}}\delta\left(x-n\frac{2\pi}{c}\right)$, for all non zero $c\in{\Bbb R}$.

\vspace{2mm}

In the above model we see that bath is essentially characterized by three functions: $f(k)$, $\omega(k)$ and $n_b(k)$. In particular,
 in what we have done so far, we have taken $f(k)=1$, $\omega(k)=\omega_b\,k$ and $n_b(k)=n_b$. This choice can be interpreted as follows:
the bath is {\em almost homogeneous}, meaning that the functions $f(k)$ and $n_b(k)$, which in principle could depend on $k$ (remember that each $k$ labels an element of the bath), are constant in $k$: different {\em members} of the bath all share
the same values of $f(k)$ and of $n_b(k)$. However, these members are not all completely identical, since $\omega(k)$ depends on $k$, as we have already observed in section \ref{DM7}.

\subsection{Changing reservoir}
\label{CR7}

To keep the situation under control as much as possible, we now discuss what happens if we keep $f(k)$ and $\omega(k)$ fixed as before, but now allow for a different dependence of $n_b(k)$ on $k$. In other words, we do not assume that $n_b(k)$ is constant in $k$, while we still take $f(k)=1$ and $\omega(k)=\omega_b\,k$. With this choice, $a(t)$ is given again as in (\ref{sr6}), and we conclude, in the same way, that $[a(t),a^\dagger(t)]=\1$. The function $n_a(t)$  is the one in (\ref{a2}) but, this time, we cannot use (\ref{a1}) since $n_b(k)$ is no longer a constant function. The computation of $\int_{\Bbb R}n_b(k)|\eta(k,t)|^2dk$ is quite similar to that of $\int_{\Bbb R}|\eta(k,t)|^2dk$, with the obvious difference due to the presence of $n_b(k)$ inside the integral. What is essential for us is that the integral is real-valued for all real choices of $n_b(k)$.  To use again the complex integration techniques, it is useful to assume that $n_b(k)$ is analytic in $k$, and that $|n_b(k)|$ does not diverge when $|k|$ diverges. For concreteness, we take
\be
n_b(k)=\frac{n_b}{k^2+\alpha^2},
\label{13}\en
for some positive $n_b$ and for some $\alpha>0$. Notice that we could safely take $\alpha<0$, too; our choice is fixed only for demonstration. On the other hand, the condition $n_b>0$ is essential because it follows from
the origin of $n_b(k)$, which came from a positive operator. The computation of $\int_{\Bbb R}n_b(k)|\eta(k,t)|^2dk$ is performed again with the techniques discussed
in details in appendix 2, with the only difference that we have two singularities of the integrating functions both in the upper and in the lower complex semi-planes, so the result of the integral is the sum of two residues.

The computation can be performed thoroughly, and an analytic form of $n_a(t)$ could be given for all $t$. However, what is more relevant for us is large time limit of $n_a(t)$, which turns out to be
\be
n_a(\infty):=\lim_{t,\infty}n_a(t)=\frac{n_b(\alpha\omega_b^2+\pi\lambda^2)}{\alpha\lambda^2\omega_b^2\left(\frac{\omega^2}{\omega_b^2}+\left(\alpha+\frac{\pi\lambda^2}{\omega_b^2}\right)^2\right)}.
\label{13bis}\en
Notice that this result makes sense only if $\alpha$ and $\lambda$ are not zero. This suggests that the complex pole in $n_b(k)$, and the interaction between Alice and her bath, are really essential to get some stabilization. We observe also that the asymptotic value of the decision function $n_a(t)$ strongly depends on the various parameters of the model, which could be adjusted to fit experimental data, if needed.

Notice that the $n_b(k)$ in (\ref{13}) describes a bath which is far from being uniform. Here, in particular, $n_b(k)\rightarrow 0$ for $|k|\rightarrow\infty$. Hence not all the parts of the bath interact with Alice in the same way: serious differences arise.

\vspace{2mm}

{\bf Remark 2.} In this section we have considered a bath which the amplitude $n_b(k)$
goes to zero as $k^{-2}$, see (\ref{13}). In fact, we could think of a function $n_b(k)$ decreasing like $k^{-1}$, but this is not compatible with the reality of the function $n_b(k)$, at least if we still want to use complex integration techniques to compute $n_a(t)$. The reason is that, to avoid singularities in the real axis, we are forced to choose $n_b(k)=\frac{\mu}{k-i\alpha}$,
for some complex $\mu$ and for some $\alpha>0$. This produces $n_a(\infty)=\gamma\omega_b$, at least if $\mu$ is chosen as follows: $\mu=\gamma\left(\omega-i\left(\alpha\omega_b+\frac{\pi \lambda^2}{\omega_b}\right)\right)$, for some real $\gamma$. So we see that, with this peculiar choice of $n_b(k)$, we still get a real $n_a(\infty)$, but what we cannot exclude that for finite time $n_a(t)$ can be complex.

\vspace{2mm}

{\bf Remark 3.} Both (\ref{sr7}) and (\ref{13bis}) suggest that the asymptotic limit of $n_a(t)$ does not really depend on whether, at $t=0$, the system is in a pure state or in a combination of states. This was already anticipated in section \ref{DM7}. We are using the Heisenberg representation. This means that the state does not change, while the observables depend on time.

\subsubsection{A remark on $f(k)$ and $\omega(k)$}

So far we have fixed $f(k)=1$ and $\omega(k)=\omega_b\,k$. This is not the only natural choice, in particular if we want to stress the difference between different parts of the reservoir. What is technically useful for us, is that Dirac's delta function appears when deducing the differential
equation for $a(t)$, see appendix 2. In fact, this is also possible under other conditions. Suppose that the two functions $f(k)$ and $\omega(k)$ are such that:
$$
\omega(k)\rightarrow \pm\infty \mbox{ for } k\rightarrow \pm\infty, \qquad \mbox { and } \qquad \frac{d\omega(k)}{dk}=\frac{1}{\beta}f^2(k),
$$
for some real $\beta\neq0$. These assumptions are satisfied if $f(k)=1$ and $\omega(k)=\omega_b k$, but also in many other situations. Hence
$$
\int_{\Bbb R}f^2(k)e^{-i\omega(k)(t-t_1)}\,dk=\beta\int_{\Bbb R}e^{-i\omega(k)(t-t_1)}\,d\omega(k)=2\pi\beta\delta(t-t_1),
$$
and we can significantly  simplify the equation for $a(t)$, which now becomes
\be
\dot a(t)=-\left(i\omega+\beta\pi\lambda^2\right) a(t)-i\lambda \int_{\Bbb{R}} f(k) b(k)\,e^{-i\omega(k)t}\,dk.
\label{14}\en
This is the equation which replaces (\ref{12}) in this new situation. The solution can be found as before, and we get
\be
a(t)=\left(a-i\lambda
\int_{\Bbb{R}}dk\,\tilde\eta(k,t)b(k)f(k)\right)e^{-\left(i\omega+\beta\pi\lambda^2\right)t},
\label{15}\en
where $\tilde\eta(k,t)=\frac{1}{\tilde\rho(k)}\left(e^{\tilde\rho(k)t}-1\right)$
and $\tilde\rho(k)=i(\omega-\omega(k))+\beta\pi\lambda^2$. At a first sight, the situation is not particularly different from those discussed earlier
and in appendix 2. However, a serious technical problem now arises: to find $[a(t),a^\dagger(t)]$ and $n_a(t)$, we need to compute integrals and the poles of the
integrating functions are strongly dependent on the analytic expression of $\omega(k)$: in particular,  if $\omega(k)$ is not linear, the computations become rather hard!

\subsection{What if we use CAR?}
\label{FREM}

In the Hamiltonian (\ref{sr4}) the operators $a$ and $b(k)$ are assumed to satisfy CCR.  However, from the point of view of a DM procedure, it might be more interesting to
consider the case in which Alice's decisions are described by CAR-operators. This means that, first of all $\{a,a^\dagger\}=a\,a^\dagger+a^\dagger a=\1$, with $\{a,a\}=0$. It is then natural to
consider a  CAR-bath as well, i.e. to assume that the operator $b(k)$ satisfies the following rules:
$$
\{b(k),b^\dagger(q)\}=\delta(k-q)\,\1, \qquad \{b(k),b(q)\}=0, \qquad \{a^\sharp,b^\sharp(k)\}=0,
$$
$x^\sharp$ being either $x$ or $x^\dagger$. Assuming the same Hamiltonian (\ref{sr4}), differential equations of motion for the annihilation operators $a(t)$ and $b(k,t)$ can be deduced. They turn out to be the same as in (\ref{sr5}), except that $\lambda$ must be replaced by $-\lambda$. In other words, we get
\be \left\{
\begin{array}{ll}
\dot a(t)=i[H,a(t)]=-i\omega a(t)+i\lambda \int_{\Bbb{R}} f(k)\,b(k,t)\,dk,\\
\dot b(k,t)=i[H,b(k,t)]=-i\omega(k) b(k,t)+i\lambda f(k)\,a(t).
\end{array}
\right. \label{eqferm} \en
As in the CCR-case, we assume that in (\ref{NB}) the function $n_b(k)$ is constant, $n_b(k)\equiv n_b \geq0.$ The CAR-setting implies that this constant cannot exceed one,
$0 \leq n_b \leq 1$ (this is the average of the CAR-number operator).

As  we can see from (\ref{sr7}), for the almost homogeneous reservoir, that $n_a(t)$ depends on $\lambda^2$ and not on $\lambda$.
The same is true also for the choice (\ref{13}) of the reservoir. This is evident from (\ref{13bis}) for large values of $t$, but can also be checked for any finite $t$. In other words, even with this different choice of $n_b(k)$, we observe that $n_a(t)$ depends on $\lambda^2$ and not on $\lambda$. The conclusion,
therefore, is clear: the CCR-CAR nature of our model does not affect the main results we have deduced. One choice or the other should be related to the nature of the decision we want to model: in case of a binary question (yes or not), the
 natural settings is probably the CAR-one. But if we assume that several possible answers (normally, infinite!) are possible, then we should use a CCR-version of the model. In this case,
the existence of a quadratic integral of motion can limit the number of possible answers to the original questions, giving rise to a sort of
finite-dimensional Hilbert space, similarly to what is discussed in details in \cite{lovestory}.

\section{Concluding discussion}
\label{conclrem}

In modeling of the process of decision making, the main output of this paper is the description of properties of an information reservoir (environment) ${\cal R}$ leading to
asymptotic stabilization. In other words, this is a description of a context  surrounding an agent, Alice, which guaranties that she would make some decision, i.e., her mind would
not fluctuate for ever between a variety of alternative answers to a question (solutions of a problem) $A.$

The presented results on asymptotic stabilization can be used to support mathematically the model of decision making as
decoherence - in its quantum field version. Here the theory of open quantum systems is applied straightforwardly
in the Hamiltonian framework. Decision making dynamics is given by the Heisenberg equations for operators describing
reflections of an agent, Alice, and the operators representing the information reservoir.
The most interesting for decision making are constraints imposed on information reservoirs, environments, guaranteeing  asymptotic stabilization.
Our present study is only a first step in this direction; further studies of sufficient and necessary conditions of asymptotic stabilization are needed.\footnote{
We remark  that applications of quantum field formalism  to modeling of decision making generate new dynamical models based on algebras of
{\it qubit creation-annihilation operators} \cite{PK}. These operators satisfy the so called qubit commutation relations which combine both
CAR and CCR-features. To the best of our knowledge,  there are no results concerning asymptotic stabilization for such dynamical processes
and this is an interesting topic for research. }
We hope that this paper will attract attention of experts in mathematical physics (especially working on mathematical problems of quantum field
theory) to the problem of asymptotic stabilization. Novel applications to social science, microeconomics, and finance, generate new laws
for functions $f(k), n_b(k), \omega(k)$ which have not been considered in physical applications.

To extend the readership of this paper, the calculations in section \ref{sect_SP}  were done at the ``physical level of rigor''.
A complete mathematical treatment of the problem should involve consideration of operator-valued distributions (generalized functions).
Such treatment definitely must be performed at some stage of future development of quantum field inspired models of decision making.

In general, the great success of quantum field theory of physics rises the expectations that this formalism will fruitfully contribute to modeling
of decision making and cognitive processes.

\section*{Appendix 1: Second quantization: CAR}
\label{AppendixA}

We briefly review some basic facts on  the so--called \emph{canonical anti-commutation
relations} (CAR), which were originally introduced in connection with what in quantum physics is called {\em second quantization} for identical particles with half-integer spin. We say that a set
of operators $\{a_\ell,\,a_\ell^\dagger, \ell=1,2,\ldots,L\}$ satisfy the CAR if the conditions
 \be \{a_\ell,a_n^\dagger\}=\delta_{\ell n}\1,\hspace{8mm} \{a_\ell,a_n\}=\{a_\ell^\dagger,a_n^\dagger\}=0 \label{a3} \en hold true for all $\ell,n=1,2,\ldots,L$. Here, $\1$ is the
identity operator and $\{x,y\}:=xy+yx$ is the \emph{anticommutator} of $x$ and $y$. These operators are those which are used to describe $L$ different \emph{modes} of fermions. From
these operators we can construct $\hat n_\ell=a_\ell^\dagger a_\ell$ and $\hat N=\sum_{\ell=1}^L \hat n_\ell$, which are both self--adjoint. In
particular, $\hat n_\ell$ is the \emph{number operator} for the $\ell$--th mode, while $\hat N$ is the \emph{number operator of $\Sc$}.
Compared with bosonic operators, the operators introduced here satisfy a very important feature: if we try to square them (or to rise to higher
powers), we simply get zero: for instance, from (\ref{a3}), we have $a_{\ell}^2=0$. This is related to the fact that fermions satisfy the Fermi
exclusion principle.

The Hilbert space of our system is constructed as follows: we introduce the \emph{vacuum} of the theory, that is a vector $\varphi_{\mathbf{0}}$
which is annihilated by all the operators $a_\ell$: $a_\ell\varphi_\mathbf{0}=0$ for all $\ell=1,2,\ldots,L$. Then we act on $\varphi_\mathbf{0}$
with the  operators $a_\ell^\dagger$ (but not higher powers, since these powers are simply zero!):
\be
\varphi_{n_1,n_2,\ldots,n_L}:=(a_1^\dagger)^{n_1}(a_2^\dagger)^{n_2}\cdots (a_L^\dagger)^{n_L}\varphi_\mathbf{0}, \label{a4} \en $n_\ell=0,1$ for
all $\ell$. These vectors give an orthonormal set and are eigenstates of both $\hat n_\ell$ and $\hat N$: $\hat
n_\ell\varphi_{n_1,n_2,\ldots,n_L}=n_\ell\varphi_{n_1,n_2,\ldots,n_L}$ and $\hat N\varphi_{n_1,n_2,\ldots,n_L}=N\varphi_{n_1,n_2,\ldots,n_L}$,
where $N=\sum_{\ell=1}^Ln_\ell$. Moreover, using the  CAR, we deduce that
\[
\hat n_\ell\left(a_\ell\varphi_{n_1,n_2,\ldots,n_L}\right)=(n_\ell-1)(a_\ell\varphi_{n_1,n_2,\ldots,n_L})
\]
and
\[
\hat n_\ell\left(a_\ell^\dagger\varphi_{n_1,n_2,\ldots,n_L}\right)=(n_\ell+1)(a_l^\dagger\varphi_{n_1,n_2,\ldots,n_L}),
\]
for all $\ell$. Then $a_\ell$ and $a_\ell^\dagger$ are  called the
\emph{annihilation} and the \emph{creation} operators. In fact, they transform the vector $\varphi_{n_1,n_2,\ldots,n_\ell,\ldots,n_L}$ into a different vector, proportional to $\varphi_{n_1,n_2,\ldots,n_\ell\pm1,\ldots,n_L}$, annihilating or creating one particle in the $\ell$-th mode. However, in some sense, $a_\ell^\dagger$ is \textbf{also} an annihilation operator since,
acting on a state with $n_\ell=1$, we destroy that state.

The Hilbert space $\Hil$ is obtained by taking  the linear span of all these vectors. Of course, $\Hil$ has a finite dimension. In particular,
for just one mode of fermions, $dim(\Hil)=2$. This also implies that, contrarily to what happens for bosons, the fermionic operators are
bounded.

The vector $\varphi_{n_1,n_2,\ldots,n_L}$ in (\ref{a4}) defines a \emph{vector (or number) state } over the algebra $\A$  as \be
\omega_{n_1,n_2,\ldots,n_L}(X)= \langle\varphi_{n_1,n_2,\ldots,n_L},X\varphi_{n_1,n_2,\ldots,n_L}\rangle, \label{a5} \en where
$\langle\,,\,\rangle$ is the scalar product in  $\Hil$.

\section*{Appendix 2: Explicit computations}
\label{Appendix}

In this appendix we give some details of our computations just sketched in Section \ref{sectAahr}, useful for those who are not familiar with this kind of interacting open systems.

Our starting point is $b(k,t)$ in (\ref{11}), with $f(k)=1$ and $\omega(k)=\omega_b\,k$. Hence
$$
b(k,t)=b(k)e^{-i\omega_bkt}-i\lambda \int_0^t a(t_1)e^{-i\omega_bk(t-t_1)}\,dt_1.
$$
When we replace this expression in the equation for $a(t)$, $\dot a(t)=-i\omega a(t)-i\lambda \int_{\Bbb{R}} b(k,t)\,dk$, we have to compute, in particular, the double integral
$$
\int_{\Bbb{R}} \left( \int_0^t a(t_1)e^{-i\omega_bk(t-t_1)}\,dt_1\right)\,dk=\int_0^ta(t_1)\left(\int_{\Bbb{R}}e^{-i\omega_bk(t-t_1)}\,dk\right)dt_1=$$
$$=\frac{2\pi}{\omega_b}\int_0^t
a(t_1)\delta(t-t_1)dt_1=\frac{\pi}{\omega_b}\,a(t),
$$
and (\ref{12}) now follows. An easy way to solve (\ref{12}) consists in using the change of variable $a(t)=A(t)\exp\left\{-\left(i\omega+\frac{\pi\lambda^2}{\omega_b}\right)t\right\}$ since, in terms of $A(t)$, (\ref{12}) can be rewritten as
$$
\dot A(t)=-i\lambda \int_{\Bbb{R}} b(k)\,e^{-i\omega_bkt}\,dk, \qquad \Rightarrow \qquad A(t)=A(0)-i\lambda \int_0^t\left(\int_{\Bbb{R}} b(k)\,e^{-i\omega_bkt_1}\,dk\right)dt_1,
$$
which returns, after few simple computation, the solution in (\ref{sr6}).

To check now that the CCR are preserved during time evolution, i.e. that $[a(t),a^\dagger(t)]=\1$ for all $t$, we observe that the operators at the right hand side of (\ref{sr6}) are the initial ($t=0$) operators, so that they satisfy the commutation rules $[a,a^\dagger]=\1$, $[b(k),b^\dagger(q)]=\delta(k-q)$, while $[a,b(k)]=[a^\dagger,b(k)]=...=0$. Therefore
$$
[a(t),a^\dagger(t)]=e^{-\frac{2\pi\lambda^2}{\omega_b}\,t}\left(1+\lambda^2\int_{\Bbb R}|\eta(k,t)|^2dk\right)\1.
$$
In our case, after some  simple algebra, we deduce that
$$
\int_{\Bbb R}|\eta(k,t)|^2dk=\int_{\Bbb R}\frac{dk}{\omega_b^2(k-k_+)(k-k_-)}\left[\left(e^{\frac{2\pi\lambda^2}{\omega_b}\,t}+1\right)-e^{(-i\omega+\frac{\pi\lambda^2}{\omega_b})t}e^{i\omega_bkt}-
e^{(i\omega+\frac{\pi\lambda^2}{\omega_b})t}e^{-i\omega_bkt}\right],
$$
where $k_\pm=\frac{\omega}{\omega_b}\pm i \frac{\pi\lambda^2}{\omega_b^2}$. Using complex integration we can compute each contribution here and we conclude that
\be
\int_{\Bbb R}|\eta(k,t)|^2dk=\frac{1}{\lambda^2}\left(e^{\frac{2\pi\lambda^2}{\omega_b}\,t}-1\right),
\label{a1}\en
so that our claim easily follows.

To prove now (\ref{sr7}), we first observe that the mean values of contributions like $a^\dagger b(k)$ and $a b^\dagger(k)$ on our states are zero. Therefore, using (\ref{sr6}),
we get
$$
n_a(t)=e^{-\frac{2\pi\lambda^2}{\omega_b}\,t}\left(\left<\varphi_{n_a},a^\dagger a\varphi_{n_a}\right>\,\left<\,
\1_\Rc\right>_\Rc+\lambda^2\left<\varphi_{n_a},\1 \varphi_{n_a}\right>\int_{\Bbb R}\int_{\Bbb R}\overline{\eta(k,t)}\eta(q,t)\,\left<\,
b^\dagger(k)b(q)\right>_\Rc\,dk\,dq\right)=
$$
\be
=e^{-\frac{2\pi\lambda^2}{\omega_b}\,t}\left(n_a+\lambda^2\int_{\Bbb R}n_b(k)|\eta(k,t)|^2dk\right).
\label{a2}\en
Then, if $n_b(k)=n_b$, we can use the result in (\ref{a1}) and recover the result in (\ref{sr7}).

\vspace{3mm}

{\bf Acknowledgments.} This paper is based on the discussions between the coauthors during the visit of two of them
(IB and KHR) to the university of Palermo in April 2017.
 IB was supported by Marie Curie Fellowship at  City University of London, H2020-MSA-IF-2015, grant N 696331; AKH was supported by
the EU-project Quantum Information Access and Retrieval Theory (QUARTZ), Grant No. 721321. FB thanks the Gruppo Nazionale di Fisica Matematica of I.N.d.A.M., and by the University of
Palermo, for support. F.B. also thanks the Distinguished Visitor Program
 of the Faculty of Science of the University of Cape Town, 2017.

\end{document}